\documentclass[pre,twocolumn,superscriptaddress]{revtex4}

\usepackage{graphicx}
\usepackage{dcolumn}

\usepackage{eucal}
\usepackage[dvips]{epsfig}
\usepackage{amssymb}

\usepackage{amsmath,amsthm}
\newcommand{\be}{\begin{eqnarray}}
\newcommand{\ee}{\end{eqnarray}}
\newcommand{\bse}{\begin{subequations}}
\newcommand{\ese}{\end{subequations}}

\newcommand{\bnum}{\begin{enumerate}}
\newcommand{\enum}{\end{enumerate}}

\newcommand{\bit}{\begin{itemize}}
\newcommand{\eit}{\end{itemize}}

\newcommand{\bc}{\begin{cases}}
\newcommand{\ec}{\end{cases}}

\newcommand{\bpm}{\begin{pmatrix}}
\newcommand{\epm}{\end{pmatrix}}

\newcommand{\bvm}{\begin{vmatrix}}
\newcommand{\evm}{\end{vmatrix}}

\newcommand{\gd}{\delta}
\newcommand{\eps}{\epsilon}%\ge schon vergeben

\newcommand{\go}{\omega}

\newcommand{\gr}{\rho}

\newcommand{\Go}{\Omega}

\newcommand{\p}{\partial}
\newcommand{\f}{\frac}

\newcommand{\lan}{\langle}
\newcommand{\ran}{\rangle}

\newcommand{\kB}{k_B}

\newcommand{\csp}{\;,\qquad}

\newcommand{\Tr}{\mathrm{Tr}\,}

\begin{document}
\title{Reply to Frenkel and Warren [arXiv:1403.4299v1]}

\author{J\"orn Dunkel}
%\email{dunkel@mit.edu}
\affiliation{Department of Mathematics, Massachusetts Institute of Technology, 
77 Massachusetts Avenue E17-412, 
Cambridge, MA 02139-4307, USA }

\author{Stefan Hilbert}
\affiliation{Max Planck Institute for Astrophysics, Karl-Schwarzschild-Str. 1, 85748 Garching, Germany}

\date{\today}
\begin{abstract} %-abstract limit 100 words, appears online only
In their paper [arXiv:1403.4299v1], Frenkel and Warren claim that the Gibbs temperature does not characterize thermal equilibrium correctly. We point out the main logical errors in their argument.
\end{abstract}

\pacs{ }
\maketitle
%%%%%%%%%%%%%%%%%%%%%%%%%%%%%%%%%%%%%%%

Frenkel and Warren (FW) criticize our paper~\cite{2014DuHa} by making the false claim   
that the Gibbs temperature does not correctly characterize the thermal equilibrium 
between two bodies.  FW's claim contradicts \emph{exact} mathematical results, as can be readily seen by considering arbitrary confined classical 
systems with conserved Hamiltonian $H(\zeta, A)=E$, where $\zeta=(\zeta_1,\ldots,\zeta_N)$ are the canonical coordinates and $A=(A_\mu)$ control parameters.  

\paragraph*{Definitions.}
We assume that $H$ is bounded from below,
$\min_\zeta H=0$, and define the microcanonical density operator $\gr_M(\zeta|E,A)={\gd(H-E)}/{\go}$, the density of states (DOS) $\go(E,A)=\Tr[\gd(E-H)]$, and the integrated DOS $\Go(E,A)=\Tr[\Theta(E-H)]$.
%\bse
%\be
%\gr_M(\zeta|E,A)&=&{\gd(H-E)}/{\go}
%\\
%\go(E,A)&=&\Tr[\gd(E-H)]
%\\
%\Go(E,A)&=&\Tr[\Theta(E-H)]
%\ee
%\ese
$\Tr$ abbreviates the phase space integral $\int d\zeta$. Adopting units $\kB=1$, the Boltzmann
entropy $S_B$ and the Gibbs entropy $S_G$ are given by 
\bse
\be
S_B&=&\ln \eps\go
\csp
T_B=\left({\p S_B}/{\p E}\right)^{-1}
\\
S_G&=&\ln \Go
\csp\;
T_G=\left({\p S_G}/{\p E}\right)^{-1}
\ee
\ese
with some energy constant $\eps$. The canonical density operator  is given by $\gr_C(\zeta|T,A)={e^{-H/T}}/{\Tr[e^{-H/T}]}$ and the Shannon entropy by $S_S=- \Tr [\gr_C \ln\gr_C]$.
% is defined by 
%\bse
%\be
%\gr_C(\zeta|T,A)&=&\f{e^{-H/T}}{\Tr[e^{-H/T}]}
%\\
%S_S&=&- \Tr [\gr_C \ln\gr_C].
%\ee
%\ese
In our paper~\cite{2014DuHa}, we formulated the requirement that a consistent thermostatistical model $(\gr, S)$ must satisfy 
\be\label{e:consistency_relation}
T \f{\p S}{\p A_\mu}=-\left \lan\f{\p H}{\p A_\mu} \right\ran_\gr 
\ee
where $\lan f\ran :=\Tr[\gr f]$ for some arbitrary function $f(\zeta)$. FW accept~\eqref{e:consistency_relation} as a valid requirement, and a considerable part of their argument builds on this criterion.

%\pagebreak
\paragraph*{Rigorous facts.}
Given the above definitions, the following three statements are mathematically~\emph{exact} results
%\vspace{0.2cm}
\begin{itemize}
\item[E1:\;] $T_G$ satisfies the microcanonical equipartition theorem
\be\label{e:equi}
\left \lan \zeta_i \f{\p H }{\p \zeta_i}\right\ran_{\gr_M} = T_G
\qquad \forall i=1,\ldots, N,
\ee
for all $N\ge 1$, whereas $T_B$ does not.

\item[E2:\;] The pair $(\gr_M,S_G)$ satisfies the consistency relation~\eqref{e:consistency_relation} for all $N\ge 1$, whereas $(\gr_M,S_B)$ does not.   
\item[E3:\;]
The pair $(\gr_C,S_S)$ satisfies the consistency relation~\eqref{e:consistency_relation} for all $N\ge 1$.   
\end{itemize}

The  proofs of E1, E2 and E3 are trivial, each taking only a few lines. 
E1 is proven in Ref.~\cite{Khinchin},  E2 in Ref.~\cite{2014DuHa}, and E3 in Ref.~\cite{2007Campisi}. 
E1, E2 and E3 suffice to invalidate the erroneous claims by FW.

\paragraph*{Incorrect claims by FW.}
FW rediscover E3 in the Appendix of their paper~\footnote{Equation~\eqref{e:consistency_relation} is equivalent to their Eq. (22)}, although the validity of E3 was already explicitly mentioned  
on page 7 in the SI of our paper~\cite{2014DuHa}. Furthermore, FW interpret their finding as evidence against ~E2.  This is logically incorrect, as E2 and E3 are unrelated and hold independently from each other.  Taken together, E2 and E3 merely imply that there exist at least two density operators $\gr$ that give rise to consistent thermostatistical models.  More importantly, however, the fact that $(\gr_C, S_S)$ and  $(\gr_M, S_G)$ are singled out by the \emph{same} criterion~\eqref{e:consistency_relation} also means that, if one accepts the Shannon entropy $S_S$ as the thermodynamic entropy of the canonical ensemble, then one must also accept the Gibbs entropy $S_G$ as the thermodynamic entropy of the microcanonical ensemble.

\par
We next address the main statement by FW, namely that $T_G$ derived from $S_G$ does not correctly characterize thermal equilibrium between two bodies. E1 immediately invalidates this claim. To demonstrate this in detail, consider 
two isolated systems 1 and  2 with canonical coordinates $z=(z_1,\ldots, z_{N_1})$ and $Z=(Z_1,\ldots, Z_{N_2})$ that are brought into thermal contact. Their joint Hamiltonian is given by 
\be
H(\zeta)=H_1(z)+H_2(Z)+\varepsilon H_{12}(z,Z),
\ee  
with $H_{12}$ denoting the interaction part, $\zeta=(z,Z)$ nd $N=N_1+N_2$.  Assume the systems had energies $E_1$ and $E_2$ before coupling. Then their total energy after coupling is $E=E_1+E_2$ and their joint microcanonical density operator is ${\gr_{12}}\propto\gd(E-H)$.  Considering the weak-coupling limit~\mbox{$\varepsilon\searrow 0$}, E1 ensures that each of the subsystems has the same temperature given by $T_G$;
explicitly, 
$$
\left \lan \zeta_i \f{\p H }{\p \zeta_i}\right\ran_{\gr_{12}} = \left \lan z_j \f{\p H_1 }{\p z_j}\right\ran_{\gr_{12}} =  \left \lan Z_k \f{\p H_2 }{\p Z_k}\right\ran_{\gr_{12}} = T_G
$$ 
%\be
%\left \lan z_i \f{\p H }{\p z_i}\right\ran_{\gr_{12}} = \left \lan z_i \f{\p H_1 }{\p z_i}\right\ran_{\gr_{12}} = T_G
%\ee 
%and for subsystem 2 
%\be
%\left \lan Z_i \f{\p H }{\p Z_i}\right\ran_{\gr_{12}} = \left \lan Z_i \f{\p H_2 }{\p Z_i}\right\ran_{\gr_{12}} = T_G.
%\ee 
Note that E1 not just guarantees that $T_G$ is \emph{the} equilibrium temperature -- E1 also implies directly that $T_B$ does not correctly characterize thermal equilibrium for any finite $N$~\footnote{It is sometimes incorrectly stated that   the energy per particle  becomes equally distributed. E1 shows that this is in general not the case; equipartition holds for the virial quantities $\zeta_i \f{\p H }{\p \zeta_i}$.  }.
\par
Last but not least, since the rigorous statements~\eqref{e:equi}, E2 and E3 are valid for arbitrary $N\ge 1$, they should also remain valid for any sensibly defined thermodynamic limit (TDL). TDLs that violate this basic continuity requirement are unsound~\footnote{If E2 and E3 yield different results in the TDL, as is the case for systems with bounded spectrum, then this just signals ensemble~\emph{inequivalence}. In such situations, one has to decide whether the underlying experiment realizes a canonical ensemble (coupling to an infinite heat bath) or a microcanonical ensemble (isolated systems with conserved energy). To our knowledge, all population-inversion experiments on spin systems or ultra-cold gases have been performed on strictly isolated systems. We also note that, since E1, E2 and E3 are valid for any finite $N\ge 1$, it is in principle unnecessary to invoke TDL arguments that require  infinite systems. The main mathematical benefit of TDLs is that phase transitions can be identified with \lq sharp\rq{} singularities when letting $N\to \infty$. For finite systems with $N<\infty$, fluctuations of thermodynamic observables can become significant and hence should also be analyzed in those cases,  but this does not affect the fact that the thermodynamic relations  E1, E2 and E3 remain true even for very small systems, provided their dynamics is sufficiently chaotic (\lq mixing\rq).}. 
\par
In summary, FW's criticism of the Gibbs temperature is unjustified~\footnote{In Eq. (5) of their paper  [arXiv:1403.4299v1], FW use a factorization approximation  that, strictly speaking, is only valid for systems with an exponentially growing DoS. Since a system with bounded spectrum typically does not exhibit an exponential growth  of the DoS near the upper band limit, the validity of FW's Eq. (5) deserves to be questioned. Another \lq hidden\rq{} assumption implicit to FW's argument is ensemble equivalence. However, for systems with bounded spectrum, microcanonical and canonical ensembles are neither mathematically nor physically equivalent; see also footnote [6] above.} and invalid~\footnote{As a matter of principle, approximate calculations cannot invalidate mathematically rigorous results like those in E1, E2 and E3.}. \newpage

\bibliographystyle{unsrt}
\bibliography{NegTemp}

\begin{thebibliography}{1}

\bibitem{2014DuHa}
J.~Dunkel and S.~Hilbert.
\newblock Consistent thermostatistics forbids negative absolute temperatures.
\newblock {\em Nature Physics}, 10:67--72, 2014.

\bibitem{Khinchin}
A.~I. Khinchin.
\newblock {\em Mathematical Foundations of Statistical Mechanics}.
\newblock Dover, New York, 1949.

\bibitem{2007Campisi}
M.~Campisi.
\newblock Thermodynamics with generalized ensembles: The class of dual
  orthodes.
\newblock {\em Physica A}, 385:501--517, 2007.

\end{thebibliography}

\end{document}